\documentclass[pra,twocolumn,preprintnumbers,amsmath,amssymb]{revtex4}
\usepackage{graphicx}
\usepackage{graphics}
\usepackage{psfrag}
\usepackage{hyperref}

\begin{document}

\title{Functional renormalization for trion formation in ultracold fermion gases}
\author{S. Floerchinger}
\author{R. Schmidt}
\author{S. Moroz}
\author{C. Wetterich}
\affiliation{Institut f\"{u}r Theoretische Physik\\Universit\"at Heidelberg\\Philosophenweg 16, D-69120 Heidelberg, Germany}

\begin{abstract}
The energy spectrum for three species of identical fermionic atoms close to a Feshbach resonance is computed by use of a nonperturbative flow equation. Already a simple truncation shows that for large scattering length $|a|$ the lowest energy state is a ``trion'' (or trimer) bound state of three atoms. At the location of the resonance, for $|a|\to\infty$, we find an infinite set of trimer bound states, with exponentially decreasing binding energy. This feature was pointed out by Efimov. It arises from limit cycle scaling, which also leads to a periodic dependence of the three body scattering coupling on $\ln |a|$. Extending our findings by continuity to nonzero density and temperature we find that a ``trion phase'' separates a BEC and a BCS phase, with interesting quantum phase transitions for $T=0$.
\end{abstract}


\maketitle

\section{Introduction}
Ultracold fermion gases with three hyperfine species near a Feshbach resonance are far from being understood as well as the analogous Bareen-Cooper-Schrieffer (BCS) - Bose-Einstein condensate (BEC) crossover with only two species \cite{ALeggett80}. Only recently such systems were investigated experimentally, showing interesting phenomena beyond the two species \cite{Ottenstein}. It is a challenge for both theory and experiment to understand the phase diagram of such systems. Although the physics of three-body scattering can be described with different methods reaching from quantum mechanics to quantum field theory, it is difficult to account for many-body effects in thermodynamic systems with a nonzero density $n$ and temperature $T$. For ultracold fermions in an optical lattice a phase with ``trions'' as well as a phase with ``color superfluidity'' was found in a variational approach \cite{Honerkamp}, see also \cite{Wilczek}.

The functional renormalization group provides a powerful method with the potential to describe both the scattering amplitudes and binding energies of the few-body sector and the many-body physics at nonzero density and temperature. A nonperturbative exact flow equation includes the effect of quantum and statistical fluctuations by following the evolution of the average action $\Gamma_k$ from ``microphysics'' to observable ``macrophysics'' \cite{Wetterich1993}. 

\begin{figure}[h!]
\centering
\includegraphics[width=\linewidth]{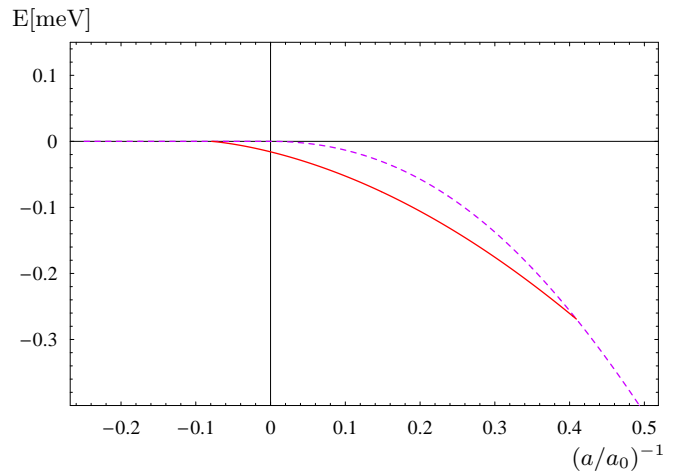}
\caption{(Color online) Energy spectrum for three fermion species close to a common Feshbach resonance. The solid line gives the energy per atom of the trion with respect to the fundamental fermion (zero energy line). The dashed line is the energy of the dimer per atom. Here $a_0$ is the Bohr radius which we also use as the ultraviolet scale $\Lambda=1/a_0$. Parameters are chosen to correspond to $^6\text{Li}$ in the ($m_F=1/2$, $m_F=-1/2$)-channel \cite{Bartenstein}.}
\label{fig:Energies}
\end{figure}

In this paper we concentrate on the functional integral description of energy levels of the dimer and trimer bound states (``vacuum physics''). We demonstrate that a rather simple truncation of the flow equation can describe the dependence of various energies on the scattering length $a$, as shown in Fig. \ref{fig:Energies}. At the ``unitarity limit'' $a\to\infty$ for $B=B_0$ we also find an infinite tower of trimer bound states, with energies given by
\begin{equation}
E_n=\exp \left(-\frac{2 \pi n}{s_0}\right)E_0,
\end{equation}
the spectrum found by Efimov in a quantum mechanical computation \cite{Efimov}, for a review see \cite{BraatenHammer}. The functional renormalization group is thus a promising tool for an investigation of the many-body physics, which can be done within the same framework of truncations, simply by changing the chemical potential and the temperature. 

Already with the present ``vacuum computation'' we can conclude by continuity that for small temperature $T$ two phases with ``extended superfluidity'' are separated by a ``trion phase'' with a different symmetry. The extended superfluidity is characterized by several Goldstone bosons, corresponding to the symmetry breaking $\text{SU(3)}\times\text{U(1)}\to \text{SU(2)}\times\text{U(1)}$ and is analogous to the BEC and BCS superfluidity \cite{He}. 

\section{Method and approximation scheme}
In a formulation with scale dependent fields, the effective average action (also called flowing action or running action) obeys the flow equation
\begin{eqnarray}
\nonumber
\partial_k \Gamma_k[\Phi_k]&=&\frac{1}{2}\text{STr}\, \left(\Gamma_k^{(2)}[\Phi_k]+R_k\right)^{-1}\partial_k R_k\\ 
&& + \left(\frac{\delta}{\delta \Phi_k}\Gamma_k[\Phi_k]\right)\partial_k \Phi_k.
\label{eq:flowequationscaledependentfields}
\end{eqnarray}
Here, the (generalized) spinor $\Phi_k$ describes all fields in the problem and $\text{STr}$ sums over momentum $\vec q$ and Matsubara frequency $q_0$ as well as the internal degrees of freedom such as species of fields, with a minus sign for fermions. By $R_k$ we denote an infrared regulator which suppresses fluctuations with momenta $\vec q^2\lesssim k^2$. At the ultraviolet scale $k=\Lambda$ the average action equals the microscopic action $\Gamma_\Lambda=S$. In our case this momentum scale is of the order of the inverse Bohr radius $\Lambda=1/a_0$, $a_0\approx5\times 10^{-11}\text{m}$. Including more and more fluctuation effects by flowing to smaller cutoff scales $k$, we finally arrive at the full quantum effective action $\Gamma=\Gamma_{k=0}$, which is the generating functional for the one-particle irreducible correlation functions. Precise knowledge of $\Gamma$ corresponds to a ``solution'' of the theory. Finding exact solutions to the flow equation \eqref{eq:flowequationscaledependentfields} is not possible in practice but one can find approximate solutions by truncating the space of functionals $\Gamma_k$ to a manageable size. Reviews on the method of functional renormalization can be found in \cite{Reviews, Pawlowski}. 

We propose a  scheme to solve Eq. \eqref{eq:flowequationscaledependentfields} approximately for the case of three species of identical fermions. Applying it to the vacuum, where $n=T=0$, yields a simple and natural description of the scattering properties. We use the following truncation for the average action
\begin{eqnarray}
\nonumber
\Gamma_k&=&\int_x {\bigg \{} \psi^\dagger\left(\partial_\tau-\Delta-\mu\right)\psi+\phi^\dagger\left(\partial_\tau-\Delta/2+m_\phi^2\right)\phi\\
\nonumber
&&+h\,\epsilon_{ijk}\,\left(\phi_i^*\psi_j\psi_k-\phi_i\psi_j^*\psi_k^*\right)/2+\lambda_\phi\left(\phi^\dagger \phi\right)^2/2\\
\nonumber
&&+\chi^*\left(\partial_\tau-\Delta/3+m_\chi^2\right)\chi+g\left(\phi_i^*\psi_i^*\chi-\phi_i\psi_i\chi^*\right)\\
&&+\lambda_{\phi\psi}\left(\phi_i^*\psi_i^*\phi_j\psi_j\right){\bigg \}}.
\label{eq:truncation}
\end{eqnarray}
Here we use natural nonrelativistic units with $\hbar=k_B=2M=1$, where $M$ is the mass of the original fermions.
The integral in Eq. \eqref{eq:truncation} goes over homogeneous space and over imaginary time as appropriate for the Matsubara formalism $\int_x=\int d^3x \int_0^{1/T}d\tau$. The symbol $\Delta$ is the Laplace operator. We denote the (Grassmann) field describing the three fermion species by $\psi=(\psi_1,\psi_2,\psi_3)$, the field of bosonic bound states by $\phi=(\phi_1,\phi_2,\phi_3)\mathrel{\widehat{=}}(\psi_2\psi_3,\psi_3\psi_1,\psi_1\psi_2)$ and also include a single component fermionic field $\chi$.  This ``trion'' field represents the totally antisymmetric combination $\psi_1 \psi_2 \psi_3$. 

On the level of the three-body sector, the symmetry of the problem would allow also for a term $\sim \psi^\dagger \psi\phi^\dagger \phi$ in Eq. \eqref{eq:truncation}. This term plays a similar role as for the case of two fermion species, where it was investigated in \cite{DKS}. We expect that the qualitative features of the three-body scattering are dominated by the term $\sim\lambda_{\phi\psi}$ in Eq. \eqref{eq:truncation}. The quantitative influence of a term $\sim \psi^\dagger \psi \phi^\dagger \phi$ on the flow equations will be investigated in future work.

We assume in this paper that the fermions $\psi_1$, $\psi_2$, and $\psi_3$ have equal mass $M$ and chemical potential $\mu$. We also assume that the interactions are independent of the spin (or hyperspin) so that our microscopic model is invariant under a global SU(3) symmetry transforming the fermion species into each other. While the fermion field $\psi=(\psi_1,\psi_2,\psi_3)$ transforms as a triplet ${\bf 3}$, the boson field $\phi=(\phi_1,\phi_2,\phi_3)$ transforms as a conjugate triplet $\bar {\bf 3}$. The trion field $\chi$ is a singlet under SU(3). In concrete experiments, for example with $^6\text{Li}$ \cite{Ottenstein}, the SU(3) symmetry may be broken explicitly since the Feshbach resonances of the different channels occur for different magnetic field values and have different widths. For nonzero density this pattern may lead to interesting phase diagrams with various types of superfluidity \cite{SuperfluidPhases}.

In addition to the SU(3) spin symmetry our model is also invariant under a global U(1) symmetry $\psi\to e^{i\alpha}\psi$, $\phi\to e^{2i\alpha}\phi$, and $\chi\to e^{3i\alpha}\chi$. The conserved charge related to this symmetry is the total particle number. Since we do not expect any anomalies the quantum effective action $\Gamma=\Gamma_{k=0}$ will also be invariant under $\text{SU(3)}\times \text{U(1)}$.

Apart from the terms quadratic in the fields that determine the propagators, Eq. \eqref{eq:truncation} contains an interaction between bosons $\sim\lambda_\phi$, a term that describes scattering between bosons and fermions $\sim\lambda_{\phi\psi}$ and the Yukawa-type interactions $\sim h$ and $\sim g$. The energy gap parameters $m_\phi^2$ for the bosons and $m_\chi^2$ for the trions are sometimes written as $m_\phi^2=\nu_\phi-2\mu$, $m_\chi^2=\nu_\chi-3\mu$, showing an explicit dependence on the chemical potential $\mu$.

In Eq. \eqref{eq:truncation}, the fermion field $\chi$ can be ``integrated out'' by inserting the $(\psi,\phi)$-dependent solution of its field equation into $\Gamma_k$. For $m_\chi^2\rightarrow \infty$ this results in a contribution to a local three-body interaction, shifting $\lambda_{\phi\psi}\rightarrow \lambda_{\phi\psi}-g^2/m_\chi^2$. Furthermore for $\lambda_\phi=0$ one may integrate out the boson field $\phi$, such that (for large $m_\phi^2$) one replaces the parts containing $\phi$ and $\chi$ in $\Gamma_k$ by an effective pointlike fermionic interaction
\begin{equation}
\Gamma_{k,\text{int}}=\int_x\frac{1}{2}\lambda_\psi(\psi^\dagger\psi)^2+\frac{1}{3!}\lambda_3 \left(\psi^\dagger \psi\right)^3,
\end{equation}
with
\begin{equation}
\lambda_\psi= -\frac{h^2}{m_\phi^2}, \quad \quad \lambda_3=\frac{h^2}{m_\phi^4} \left(\lambda_{\phi\psi}-\frac{g^2}{m_\chi^2}\right).
\label{eq:subst}
\end{equation}

We note that the contribution of trion exchange to $\lambda_{\phi\psi}$ or $\lambda_3$ depends only on the combination $g^2/m_\chi^2$. The sign of $g$ can be changed by $\chi\rightarrow - \chi$, and the sign of $g^2$ can be reversed by a sign flip of the term quadratic in $\chi$. Keeping the possible reinterpretation by this mapping in mind, we will formally also admit negative $g^2$ (imaginary $g$).

At the microscopic scale $k=\Lambda$, we use the initial values of the couplings in Eq. \eqref{eq:truncation} $g=\lambda_\phi=\lambda_{\phi\psi}=0$ and $m^2_\chi\to\infty$. Then the fermionic field $\chi$ decouples from the other fields and is only an auxiliary field which is not propagating. However, depending on the parameters of our model we will find that $\chi$, which describes a composite bound state of three original fermions $\chi=\psi_1\psi_2\psi_3$, becomes a propagating degree of freedom in the infrared. The field $\chi$ is called trion. The initial values of the boson energy gap $\nu_\phi$ and the Yukawa coupling $h$ will determine the scattering length $a$ between fermions and the width of the resonance, see below. The pointlike limit (broad resonance) corresponds to $\nu_\phi\to\infty$, $h^2\to\infty$ where the limits are taken such that the effective renormalized four fermion interaction remains fixed. In Eq. (\ref{eq:truncation}) we use renormalized fields $\phi=\bar A_\phi^{1/2}(k)\, \bar{\phi}$, $\psi=\bar A_\psi^{1/2}(k)\,\bar \psi$, $\chi = \bar A_\chi^{1/2}(k)\,\bar \chi$, with $\bar A_\phi(\Lambda)=\bar A_\psi(\Lambda)=\bar A_\chi(\Lambda)=1$, and renormalized couplings $m_\phi^2=\bar{m}_\phi^2/\bar{A}_\phi$, $h=\bar{h}/(\bar{A}_\phi^{1/2}\bar A_\psi)$, $\lambda_\phi=\bar \lambda_\phi/\bar A_\phi^2$, $m_\chi^2=\bar m_\chi^2/\bar A_\chi$, $g=\bar g /(\bar A_\chi^{1/2}\bar A_\phi^{1/2}\bar A_\chi^{1/2})$, and $\lambda_{\phi\psi}=\bar \lambda_{\phi\psi}/(\bar A_\phi \bar A_\psi)$.

To derive the flow equations for the couplings in Eq. \eqref{eq:truncation} we have to specify an infrared regulator function $R_k$. Here we use the particularly simple function $R_{k}=r(k^2-\vec p^2)\theta(k^2-\vec p^2)$, where $r=1$ for the fermions $\psi$, $r=1/2$ for the bosons $\phi$, and $r=1/3$ for the composite fermionic field $\chi$. This choice has the advantage that we can derive analytic expressions for the flow equations and that it is optimized in the sense of \cite{Litim}. We use the method of ``(re-)bosonization'' \cite{Gies2002}, neglecting correction terms which have only a minor quantitative effect.

\section{Flow equations for two-body sector}
We can now insert our ansatz equation \eqref{eq:truncation} into the flow equation \eqref{eq:flowequationscaledependentfields}. Functional derivatives with respect to the fields for zero temperature $T=0$ and density $n=0$ lead us to a system of ordinary coupled nonlinear differential equations for the couplings $\bar m_\phi^2$, $\bar h$, $\bar A_\psi$, $\bar{A}_\phi$, $\bar{A}_\chi$, $\bar m_\chi^2$, $\bar g$ and $\bar \lambda_{\phi\psi}$. One finds that the propagator of the original fermions $\psi$ is not renormalized, $\bar A_\psi(k)=1$, $\bar h(k) = \bar h$. The flow equations for the couplings determining the two body sector, namely the boson gap parameter (with $t=\ln (k/\Lambda)$)
\begin{equation}
\partial_t \bar m_\phi^2 = \frac{\bar h^2}{6\pi^2}  \frac{k^5}{(k^2-\mu)^2},
\end{equation}
and the boson wave function renormalization
\begin{equation}
\partial_t \bar A_\phi = -\frac{\bar h^2}{6\pi^2}\frac{k^5}{(k^2-\mu)^3}
\end{equation}
decouple from the other flow equations. They can be solved analytically \cite{Diehl2007a}
\begin{eqnarray}
\nonumber
\bar m_\phi^2(k) &=& \bar m_\phi^2(\Lambda)-\frac{\bar h^2}{6\pi^2}{\bigg [} (\Lambda-k)-\frac{\mu}{2}\left(\frac{\Lambda}{\Lambda^2-\mu}-\frac{k}{k^2-\mu}\right)\\
\nonumber
&+&\frac{3}{2}\sqrt{-\mu}\left(\text{arctan}\left(\frac{\sqrt{-\mu}}{\Lambda}\right)-\text{arctan}\left(\frac{\sqrt{-\mu}}{k}\right)\right){\bigg ]}\\
\nonumber
\bar A_\phi(k) &=& 1
-\frac{\bar h^2}{6\pi^2}{\bigg [} \frac{\Lambda(5\Lambda^2-3\mu)}{8(\Lambda^2-\mu)^2}-\frac{k(5k^2-3\mu)}{8(k^2-\mu)^2}\\
\nonumber
&+&\frac{3}{8\sqrt{-\mu}}\left(\text{arctan}\left(\frac{\sqrt{-\mu}}{\Lambda}\right)-\text{arctan}\left(\frac{\sqrt{-\mu}}{k}\right)\right){\bigg ]}.\\
\label{eq:explicitsolutiontwobody}
\end{eqnarray}
Using this explicit solution \eqref{eq:explicitsolutiontwobody} we can relate the initial value of the boson gap parameter $\bar m_\phi^2$ as well as the Yukawa coupling $\bar h$ to physical observables. The interaction between fermions $\psi$ is mediated by the exchange of the bound state $\phi$. Again in terms of bare quantities, the scattering length between fermions is given by $a=-\bar h^2/(8\pi\bar m_\phi^2)$ where the couplings $\bar h$ and $\bar m_\phi^2$ are evaluated at the macroscopic scale $k=0$ and for vanishing chemical potential $\mu=0$ \cite{Diehl2008}. This fixes the initial value
\begin{equation}
\bar m_\phi^2(\Lambda)=-\frac{\bar h^2}{8\pi}a^{-1}+\frac{\bar h^2}{6\pi^2}\Lambda-2\mu.
\end{equation}
In addition to the $\mu$-independent part we have added here the  chemical potential term $-2\mu$ where the factor $2$ accounts for the bosons consisting of two fermions. In vacuum, the gap parameter of the bosons $\bar{m}_\phi^2$ is proportional to the detuning of the magnetic field $\bar m_\phi^2(k=0,\mu=0)=\mu_M (B-B_0)$.
Here $\mu_M$ is the magnetic moment of the bosonic dimer and $B_0$ is the magnetic field at the resonance. From
\begin{equation}
a=-\frac{1}{8\pi} \frac{\bar h^2}{\mu_M (B-B_0)}
\end{equation}
one can read off that $\bar h^2$ is proportional to the width of the resonance. We have now fixed all initial values of the couplings at the scale $k=\Lambda$ or, in other words, the parameters of our microscopic model.

At vanishing density $n=0$, the chemical potential is negative or zero, $\mu\leq0$, and will be adjusted such that the lowest excitation of the vacuum is a gapless propagating degree of freedom in the infrared, i.~e. at $k=0$. Depending on the value of $a^{-1}$ this lowest energy level may be the original fermion $\psi$, the boson $\phi$, or the composite fermion $\chi$. 

\section{Three-body problem}
Now that we have solved the equations for the two-body problem within our approximation, we can address the three-body sector. It is described by the flow equations for the trion gap parameter,
\begin{equation}
\partial_t m_\chi^2 = \frac{6 g^2}{\pi^2} \frac{k^5}{(3k^2-2\mu +2 m_\phi^2)^2}+\eta_\chi m_\chi^2,
\label{eq:flowmchi}
\end{equation}
and the Yukawa-type coupling,
\begin{eqnarray}
\partial_t g &=& (\eta_\phi+\eta_\chi)\frac{g}{2}+m_\chi^2\partial_t \alpha \notag \\ 
& & -\frac{2g h^2}{3\pi^2} \frac{k^5}{(k^2-\mu)^2}\frac{ 6k^2-5\mu + 2 m_\phi^2}{(3k^2-2\mu+2m_\phi^2)^2}.
\label{eq:flowg2}
\end{eqnarray}
These equations are supplemented by the anomalous dimension
\begin{equation}
\eta_\chi =-\frac{\partial_t \bar A_\chi}{\bar A_\chi} =\frac{24 g^2}{\pi^2} \frac{k^5}{ (3k^2-2\mu+2m_\phi^2)^3}
\label{eq:etachi}
\end{equation}
and the variable
\begin{equation}
\partial_t \alpha = -\frac{h^4}{12\pi^2 g}\frac{ k^5}{(k^2-\mu)^3}\frac{9k^2-7\mu +4m_\phi^2}{  (3k^2-2\mu+2m_\phi^2)^2}.
\label{eq:alpha}
\end{equation}
Here $\alpha$ determines the scale dependence of the trion field $\chi$ 
\begin{equation}
\partial_t \chi =(\partial_t \alpha)\phi_i\psi_i+\eta_\chi \chi/2.
\label{eq:scaledepfield}
\end{equation}
While the second term in Eq. \eqref{eq:scaledepfield} is the usual wave-function renormalization, the first term describes a nonlinear change of variables. It is chosen such that the flow of $\lambda_{\phi\psi}$ vanishes on all scales, $\lambda_{\phi\psi}(k)=0$. The scattering between bosons and fermions is then described by the exchange of the trion bound state $\chi$. This reparametrization, which is analogous to rebosonization \cite{Gies2002, Pawlowski}, is crucial for our description of the system in terms of the composite trion field $\chi$. 
Note that the flow equations \eqref{eq:flowmchi}, \eqref{eq:flowg2}, \eqref{eq:etachi}, \eqref{eq:alpha} that describe the three-body sector are independent from the flow of the boson-boson interaction $\lambda_\phi$ that belongs to the four-body sector. 

\begin{figure}[h!]
\centering
\includegraphics[width=\linewidth]{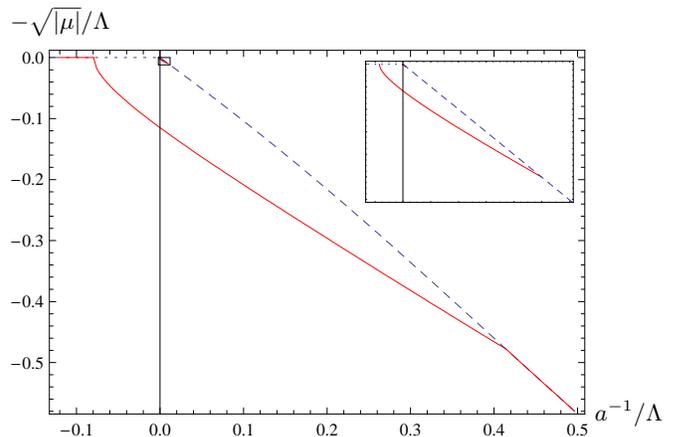}
\caption{(Color online) Dimensionless chemical potential $-\sqrt{|\mu|}/\Lambda$ in vacuum as a function of the dimensionless scattering length $a^{-1}/\Lambda$. For comparison, we also plot the result for two fermion species where the original fermions are the propagating particles for $a^{-1}<0$ (dotted) and composite bosons for $a^{-1}>0$ (dashed). The inset is a magnification of the little box and shows the energy of the first excited Efimov state.}
\label{fig:Efimov}
\end{figure}

Since the three-body sector is driven by fermionic and bosonic fluctuations, it is not possible to find simple analytic solutions to the flow equations in the general case. However, it is no problem to solve them numerically. This is most conveniently done using again ``bare'' couplings. As a general feature we note that a negative chemical potential $\mu$ acts as an infrared cutoff for the fermionic fluctuations while a positive value of the bosonic gap $m_\phi^2$ suppresses bosonic fluctuations in the infrared. We can find numerical solutions for different scattering length between the fermions $a$ and varying chemical potential $\mu$. We use an iteration process to determine the chemical potential $\mu\leq0$ where the lowest excitation of the vacuum is gapless. On the far BCS side for small and negative scattering length $a\to0_-$, where this lowest excitation is the fermion $\psi$, this implies simply $\mu=0$. On the far BEC side for small and positive scattering length $a\to0_+$ the lowest excitation is the boson $\phi$, implying $\bar m_\phi^2 =0$. We can then use our analytic solution of the two-body sector Eq. \eqref{eq:explicitsolutiontwobody} to obtain the chemical potential $\mu$ from the condition $\bar m_\phi^2=0$ at the macroscpic scale $k=0$. In the limit $\Lambda/|\mu|\to\infty$ we find 
\begin{equation}
\mu =-\left(\sqrt{\frac{\bar h^4}{(32 \pi)^2}+\frac{a^{-1}\bar h^2}{16 \pi}}-\frac{\bar h^2}{32 \pi}\right)^2.
\end{equation}
In the broad resonance limit $\bar h^2\to \infty$ this reduces to the well-known result $\mu=-1/a^2$.
For scattering lengths close to the Feshbach resonance $a^{-1}_{c1}<a^{-1}<a^{-1}_{c2}$ we find that the lowest vacuum excitation is the trion $\chi$. Numerically, we determine $\mu$ from the implicit equation $m_\chi^2=0$ at $k=0$. Our result for $\mu$ obtained for $\bar h^2=100\Lambda$ as a function of the inverse scattering length $a^{-1}$ is shown in Fig. \ref{fig:Efimov}. With the choice $\Lambda=1/a_0$, this value of $\bar h^2$ corresponds to the width of the Feshbach resonance of $^{6}\textrm{Li}$ atoms in the $(m_F=1/2,\ m_F=-1/2)$-channel \cite{Bartenstein}. 

\begin{figure}[h!]
\centering
\includegraphics[width=\linewidth]{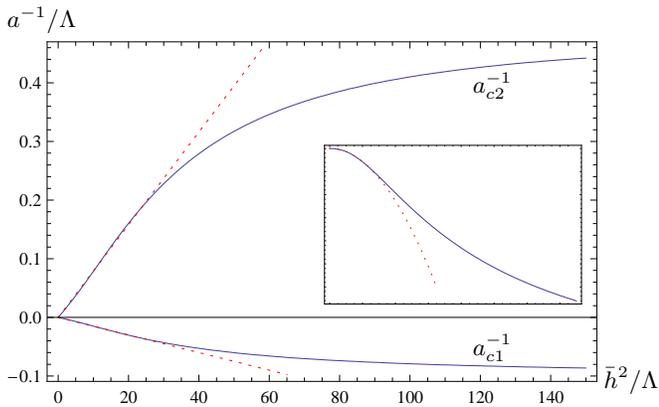}
\caption{(Color online) Interval of scattering length $a^{-1}_{c1}<a^{-1}<a^{-1}_{c2}$ where the lowest vacuum excitation is the trimer fermion $\chi$ (solid lines). For comparison we also plot the linear fits $a^{-1}_{c1}=-0.0015\,\bar h^2$ and $a^{-1}_{c2}=0.0079\,\bar h^2$ (dotted). The inset shows the chemical potential $\mu_U/\Lambda^2$ at the resonance $a^{-1}=0$ as a function of $\bar h^2/\Lambda$ in the same range of $\bar h^2/\Lambda$. Here we also plot the curve $\mu_U=-3.5\times10^{-6}\bar h^4$ for comparison (dotted).}
\label{fig:ascaling}
\end{figure}

For small Yukawa couplings $\bar h^2/\Lambda\ll 1$, or narrow resonances we find that the range of scattering length where the trimer is the lowest excitation of the vacuum increases linear with $\bar h^2$ \cite{Petrov, Mora}. More explicit, we find $a^{-1}_{c1}=-0.0015\,\bar h^2$, $a^{-1}_{c2}=0.0079\,\bar h^2$. However, for very broad Feshbach resonances $\bar h^2/\Lambda\gg1$ the range depends on the ultraviolet scale $a^{-1}_{c1},a^{-1}_{c2}\sim \Lambda$. We show this behavior in Fig. \ref{fig:ascaling}.

The chemical potential $\mu_U$ at the unitarity point $a^{-1}=0$ is plotted in the inset of Fig. \ref{fig:ascaling}. It increases with the width of the resonance similar to $\mu_U=-3.5\times 10^{-6}\bar h^4$ for small $\bar h^2/\Lambda$. It also approaches a constant value which depends on the cutoff scale $\mu_U\sim\Lambda^2$ in the broad resonance limit $\bar h^2\to \infty$. 

The dispersion relation for the atoms, dimers and trimers can be computed by analytical continuation, $\tau=it$, of the inverse propagators to ``real frequencies'' $\omega$. In our truncation, they read (in terms of renormalized fields)
\begin{eqnarray}
P_\psi&=&\frac{\vec{p}^2}{2M}-\omega_\psi-\mu,\notag\\
P_\phi&=&\frac{\vec{p}^2}{4M}-\omega_\phi-2\mu+\nu_\phi(\mu+\omega_\phi/2-\vec{p}^2/(8M)),\notag\\
P_\chi&=&\frac{\vec{p}^2}{6M}-\omega_\chi-3\mu+\nu_\chi(\mu+\omega_\chi/3-\vec{p}^{2}/(18M)).\notag\\
\label{eq:propagators}
\end{eqnarray}
We note that the functions $\nu_\phi$ and $\nu_\chi$ depend only on the particular combinations of $\mu$, $\omega$, and $\vec{p}^2$ given above. This is a result of the symmetries of our problem. The real-time microscopic action $S=\Gamma_{k=\Lambda}$ with $t=-i\tau$ is invariant under the time-dependent U(1) symmetry transformation of the fields $\psi\to e^{iEt}\psi$, $\phi\to e^{i2Et}\phi$, and $\chi\to e^{i3Et}\chi$ if we also change the chemical potential according to $\mu\to\tilde \mu=\mu+E$. Since the microscopic action has this symmetry and since we do not expect any anomalies, the quantum effective action $\Gamma=\Gamma_{k=0}$ is also invariant under these transformations. In consequence, any dependence on $\mu$ must be accompanied by a corresponding frequency dependence, such that only the invariant combinations $\omega_\psi+\mu$, $\omega_\phi+2 \mu$, $\omega_\chi+3\mu$ appear in the effective action. The relation between the dependence on $\omega$ and on $\vec{p}^2$ is a consequence of Galilean symmetry. Even though we compute directly with the flow equations only $\nu_\phi(\mu,\omega_\phi=0,\vec{p}^2=0)$ and $\nu_\chi(\mu,\omega_\chi=0,\vec{p}^2=0)$, we can use the symmetry information (``Ward identities'') for an extrapolation to arbitrary $\omega$ and $\vec{p}^2$. The dispersion relation for the bosons $\mathcal{E}_\phi(\vec{p}^2)$ follows from Eq. \eqref{eq:propagators} by solving $P_\phi(\omega_\phi=\mathcal{E}_\phi)=0$ and similarly the dispersion relation of the fermions $\mathcal{E}_\psi(\vec{p}^2)$ and the trions $\mathcal{E}_\chi(\vec{p}^2)$.

We are interested in the dimer and trimer energy levels relative to the energy of the atoms. For this purpose we consider their dispersion relations at rest ($\vec{p}^2=0$) and substract twice or three times the atom energy $\mathcal{E}_\psi=-\mu$,
\begin{eqnarray}
E_\phi&=&\mathcal{E}_\phi(\vec{p}^2=0)+2 \mu=\nu_\phi(E_\phi/2),\notag\\
E_\chi&=&\mathcal{E}_\chi(\vec p^2=0)+3\mu=\nu_\chi(E_\chi/3).
\label{eq:energylevels}
\end{eqnarray}
When the dimers are the lowest states one has $m^2_\phi=0$ and therefore $\nu_\phi=2 \mu=E_\phi$, while for lowest trimers $m_\chi^2=0$ implies $\nu_\chi=3\mu=E_\chi$. We have shown $E_\phi$ and $E_\chi$ in Fig. \ref{fig:Energies}. A typical value for $\Lambda$ is of the order of the inverse Bohr radius $a_0^{-1}$. We also mention that the energy levels have been computed in the absence of a microscopic three-body interaction, $\lambda_3(\Lambda)=0$. The same computation may be repeated with nonzero $\lambda_3(\Lambda)$, and thus microscopic parameters may be fixed by a comparison to the experimentally measured spectrum.

So far we were only concerned with the lowest energy excitation of the vacuum. We found that near a Feshbach resonance this lowest excitation is given by a trimer. However, it is known from the work of Efimov \cite{Efimov} that one can expect not only one trimer state close to resonance but a whole spectrum. In fact, the solution $\nu_\chi=3\mu$ may not be the only solution for the equation fixing the trimere energy levels, i. e.
\begin{equation}
\mathcal{E}_\chi=\nu_\chi\left(\mu+\frac{1}{3}\mathcal{E}_\chi\right)-3\mu.
\label{eq:TrionGroundEnergy}
\end{equation}
For an investigation of the energy dependence of $\nu_\chi$ we use the symmetry transformation and shift the chemical potential by $\mathcal{E}_\psi$, $\tilde{\mu}=\mu+\mathcal{E}_\psi=0$. We can therefore follow the flow at vanishing $\tilde{\mu}$ and Eq. \eqref{eq:TrionGroundEnergy} turns into a simple implicite equation for the trion energy levels
\begin{equation}
E_\chi=\nu_\chi\left(\frac{E_\chi}{3}\right)=m^2_\chi\left(\frac{E_\chi}{3}\right)
\label{eq:Echi}
\end{equation}

For $\tilde{\mu}=0$ the flow equations simplify considerably. For example, Eq. \eqref{eq:explicitsolutiontwobody} becomes
\begin{eqnarray}
\bar m^2_\phi(k)&=&\bar m^2_\phi(\Lambda)-\frac{\bar h^2}{6 \pi^2} (\Lambda-k)=\mu_M (B-B_0)+\frac{\bar h^2 k}{6 \pi^2},\notag \\
\bar A_\phi(k)&=&1+\frac{\bar h^2}{6 \pi^2}\left(\frac{1}{k}-\frac{1}{\Lambda}\right).
\end{eqnarray}
We observe that the two-body sector is evaluated here for $\tilde{\omega}_\phi=\omega_\phi+2\mu=0$. The physical chemical potential $\mu$ remains, of course, negative in the vicinity of the Feshbach resonance, and our evaluation therefore corresponds to a positive energy $\omega_\phi$ as compared to the lowest energy level of the trimer for which $\omega_\chi=0$.

An especially interesting point in the spectrum is the unitarity limit, $B=B_0$, where the scattering length diverges, $a^{-1}=0$. At that point all length scales drop out of the problem and we expect a sort of scaling solution for the flow equations. In the limit $k\to 0$ the solution for $\bar A_\phi(k)$ is dominated completely by the term with $1/k$, $\bar A_\phi(k)=\bar h^2/(6\pi^2 k)$, and we find $m_\phi^2=\bar m_\phi^2/\bar A_\phi=k^2$. For $E_\chi\to 0$ the flow equations for the three-body sector simplify and we find 
\begin{eqnarray}
\nonumber
\partial_t \bar m_\chi^2 &=& \frac{36}{25} \bar g^2\frac{k^2}{\bar h^2}\\
\partial_t \bar g^2 &=& -\frac{64}{25}\bar g^2-\frac{13}{25}\bar m_\chi^2 \frac{\bar h^2}{k^2}.
\label{eq:mgsystem}
\end{eqnarray}
In contrast, for $E_\chi\neq 0$, as needed for Eq. \eqref{eq:Echi}, we will have to solve the flow with a nonvanishing ``effective chemical potential'' $\hat{\mu}=E_\chi/3$, which will cause a departure from the scaling flow.

\section{Limit cycle scaling} 
First, let us consider the scaling solution obeying Eq. \eqref{eq:mgsystem}. It is convenient to rescale the variables according to $\tilde m^2=\bar m_\chi^2 (k/\bar h)^\theta$, $\tilde g^2= \bar g^2(k/\bar h)^{2+\theta}$ which gives the linear differential equation
\begin{equation}
\partial_t \begin{pmatrix}\tilde m^2 \\ \tilde g^2\end{pmatrix}=\begin{pmatrix}\theta, & \frac{36}{25}\\ -\frac{13}{25}, & 2+\theta-\frac{64}{25}\end{pmatrix} \begin{pmatrix}\tilde m^2 \\ \tilde g^2\end{pmatrix}.
\label{eq:matrixdifferentialequation}
\end{equation}
The matrix on the right hand side of Eq. \eqref{eq:matrixdifferentialequation} has the eigenvalues
$\beta_{1/2}=\theta-(7/25)\pm i \sqrt{419}/25$.
The flow equation \eqref{eq:matrixdifferentialequation} leads therefore to an oscillating behavior. It is straightforward to solve Eq. \eqref{eq:matrixdifferentialequation} explicitly. Restricting to real solutions and using initially $\bar g^2(\Lambda)=0$ we find the following for the trimer gap parameter and coupling
\begin{eqnarray}
\bar m_\chi^2(k)&=&\left(\frac{k}{\Lambda}\right)^{-\frac{7}{25}}\bar m_\chi^2(\Lambda)\Big[\cos\left(s_0 \ln\frac{k}{\Lambda}\right)\notag \\ 
& &  +\frac{7}{\sqrt{419}}\sin\left( s_0 \ln \frac{k}{\Lambda}\right)\Big],\notag \\
\bar g^2(k)&=&-\frac{13 \,\bar h^2}{\sqrt{419}\,k^2}\left(\frac{k}{\Lambda}\right)^{-\frac{7}{25}} \bar m_\chi^2(\Lambda) \sin\left(s_0 \ln \frac{k}{\Lambda} \right).\notag\\
\label{eq:mgsolution}
\end{eqnarray}
As it should be, the initial value $\bar m_\chi^2(\Lambda)$ drops out of the ratio $\bar g^2/\bar m^2_\chi$ and we find for the three-body coupling
\begin{equation}
\lambda_3(k)=\frac{468 \pi^4}{\sqrt{419}}\frac{\sin\left(s_0 \ln \frac{k}{\Lambda}\right)}{\cos\left( s_0\ln \frac{k}{\Lambda} \right)+\frac{7}{\sqrt{419}}\sin\left(s_0 \ln \frac{k}{\Lambda}\right)}k^{-4}.
\label{eq:lambda3}
\end{equation}
We obtain for the ``frequency'' $s_0=\sqrt{419}/25\approx 0.82$. Since we use a truncation in the space of functionals $\Gamma_k$ this result is only a rough estimate. It has to be compared with the result of other methods which find $s_0\approx1.00624$ \cite{Efimov,Bedaque,BraatenHammer}. Considered the simplicity of our approximation, the agreement is quite reasonable.

For a determination of the trion energy levels we have to solve the flow with an effective negative chemical potential $\tilde{\mu}=E_\chi/3$. This acts as an infrared cutoff, such that the flow deviates from the limit cycle once $k^2\approx-\tilde{\mu}$. Qualitatively, the flow eventually stops once $k$ becomes smaller than $\sqrt{-\tilde{\mu}}$. For an evaluation of Eq. \eqref{eq:Echi} we may therefore use Eq. \eqref{eq:mgsolution} with a specific value for $k$, namely $k^2=-E_\chi/3$. The possible energy levels therefore obey \begin{equation}
3k^2+\bar A^{-1}_\chi(k)\,\bar m_\chi^2(k)=0.
\label{eq:mchicond}
\end{equation}
With
\begin{equation}
\partial_k\bar A_\chi(k)=\frac{24}{125\pi^2} \frac{\bar g^2(k)}{k^2}
\end{equation}
one finds, up to oscillatory behavior an increase of $\bar A_\chi \sim k^{-\frac{32}{25}}$. We can write Eq. \eqref{eq:mchicond} in the form
\begin{eqnarray}
F(k)&=&\cos\left(s_0 \ln\frac{k}{\Lambda}\right)+\frac{7}{\sqrt{419}}\sin\left(s_0 \ln\frac{k}{\Lambda}\right)\notag\\
&=&-3 \frac{k^2}{\bar m_\chi^2(\Lambda)}\left(\frac{k}{\Lambda}\right)^{\frac{7}{25}}\bar A_\chi(k).
\end{eqnarray}
For small $k\ll\Lambda$ we infer that $F(k)$ has to vanish $\sim -k/ \bar h^2$. Since $F(k)$ is periodic, solutions will occur for roughly equidistant values in $\ln\frac{k}{\Lambda}$.

For $k\ll\bar h^2$ the possible solutions simply correspond to $F(k)=0$. The first solution with the largest $k$ corresponds to the ground state level with $E_0<0$. The subsequent solutions obey 
\begin{equation}
s_0\left(\ln\frac{k_{n+1}}{\Lambda}-\ln\frac{k_{n}}{\Lambda} \right)=-\pi
\end{equation}
or
\begin{equation}
\frac{E_{n+1}}{E_n}=\exp\left( -\frac{2 \pi}{s_0} \right),\quad E_n=\exp\left( -\frac{2 \pi n}{s_0} \right)E_0.
\label{eq:energyscaling}
\end{equation}
This corresponds to the tower of trimer bound states at the unitarity limit, with $E_n$ approaching zero exponentially for $n\to\infty$. For Eq. \eqref{eq:energyscaling} we have actually taken into account all zeros of $F(k)$. Since $\bar g^2(k)$ oscillates periodically, only half of these zeros correspond to $\bar g^2(k)>0$, while the other half has formally $\bar g^2(k)<0$. We may use the mapping discussed after Eq. \eqref{eq:subst} to obtain an equivalent picture with positive $\bar g^2$.

\begin{figure}[h!]
\centering
\includegraphics[width=\linewidth]{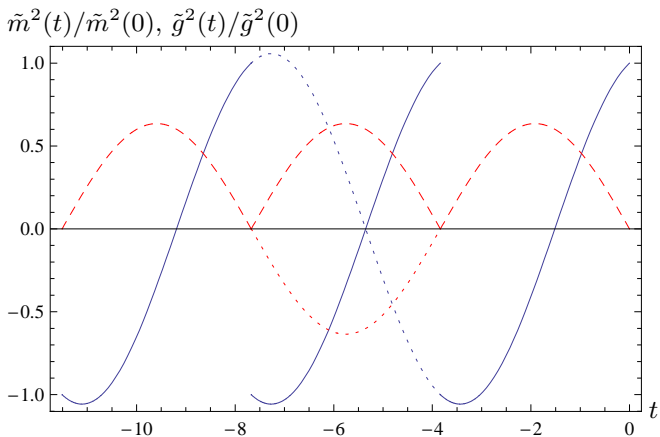}
\caption{(Color online) Limit cycle in the renormalization group flow at the unitarity point $a^{-1}=0$, and for energy at the fermion threshold $\tilde{\mu}=\mu+E=0$. We plot the rescaled gap parameter of the trimer $\tilde m^2(t)$ (solid) and the rescaled Yukawa coupling $\tilde g^2(t)$ (dashed). The dotted curves would be obtained from naive continuation of the flow after the point where $\tilde g^2=0$.}
\label{fig:limitcycla}
\end{figure}

We may understand the repetition of states by the following qualitative picture. The coupling $\tilde m^2$, that is proportional to the energy gap of the trimer, starts on the ultraviolet scale with some positive value. The precise initial value is not important. The Yukawa-type coupling $\bar g$ vanishes initially so that the trion field $\chi$ is simply an auxiliary field which decouples from the other fields and is not propagating. However, quantum fluctuations lead to the emergence of a scattering amplitude between the original fermions $\psi$ and the bosons $\phi$. We describe this by the exchange of a composite fermion $\chi$. 
This leads to an increase of the coupling $\tilde g^2$ and a decrease of the trion gap $\tilde m^2$. At some scale $t_1=\text{ln}(k_1/\Lambda)$ with $k_1^2\approx -\mu$ the coupling $\tilde m^2$ crosses zero which indicates that a trion state $\chi$ becomes the lowest energy excitation of the vacuum. Indeed, would we consider the flow without modifying the chemical potential, this would set an infrared cutoff that stops the flow at the scale $k_1\approx\sqrt{|\mu|}$ and the trion $\chi$ would be the gapless propagating particle while the original fermions $\psi$ and the bosons $\phi$ are gapped since they have higher energy. 

Following the flow further to the infrared, we find that the Yukawa coupling $\tilde g^2$ decreases again until it reaches the point $\tilde g^2=0$ at $t=t_1^\prime$ (see Fig. \ref{fig:limitcycla}). Naive continuation of the flow below that scale would lead to $\tilde g^2<0$ and therefore imaginary Yukawa coupling $\tilde g$. However, since the trion field $\chi$ decouples from the other fields for $\tilde g=0$, we are not forced to use the same field $\chi$ as before. We can simply use another auxiliary field $\chi_2$ with very large gap $m^2_{\chi_2}=\bar m_{\chi_2}^2/\bar A_{\chi_2}$ to describe the scattering between fermions and bosons on scales $t<t_1^\prime$. We are then in the same position as on the scale $k=\Lambda$ and the process repeats. Starting from a positive value, the rescaled gap parameter $\tilde m_{\chi2}^2$ decreases as the infrared cutoff $k$ is lowered. At the scale $k_2$ it crosses zero which indicates that there is a second trimer bound state in the spectrum with energy per original fermion $E_2=-\mu-k_2^2$. Would we use the modified chemical potential $\tilde \mu =\mu+E_2=-k_2^2$, the flow would be stopped at the scale $k_2$ and the second trimer $\chi_2$ would be the propagating degree of freedom. This cycle repeats and corresponds precisely to the limit cycle scaling of Eq. \eqref{eq:matrixdifferentialequation}.

At the unitarity point with $a^{-1}=0$ and at the threshold energy $E=-\mu$ with $\tilde \mu=\mu+E=0$, this limit cycle scaling is not stopped and leads to an infinite tower of trimer bound states. The energy of this states comes closer and closer to the fermion-boson threshold energy $E=-\mu$.  In our language we recover the effect first predicted by Efimov \cite{Efimov}. 
A similar limit cycle description of the Efimov effect for identical bosons was given in the context of effective field theory in \cite{Bedaque}.

The infinite limit cycle scaling occurs only directly at the Feshbach resonance with $a^{-1}=0$. Similar to a nonzero (modified) chemical potential $\tilde \mu$, also a nonzero inverse scattering length $a^{-1}$ provides an infrared cutoff that stops the flow. For example for negative scattering length $a$ and energy $E$ with $\tilde \mu=\mu+E=0$, the solution of the two body sector Eq. \eqref{eq:explicitsolutiontwobody} implies for the boson gap in the infrared $m_\phi^2=-3\pi a^{-1}k/4 + k^2$ so that bosonic fluctuations are suppressed in comparison to the unitarity point with $a^{-1}=0$. This leads to a stop of the limit cycle scaling at the scale $k\approx3\pi |a^{-1}|/4$ so that the spectrum consists only of a finite number of trimer states. In the inset of Fig. \ref{fig:Efimov} we show the numerical result for the modified chemical potential $\tilde \mu$ of the first excited Efimov trimer.

\section{Conclusion}
Our simple truncation of an exact functional renormalization group equation yields a quite detailed picture for the spectrum of bound states and the scattering of three species of identical fermions. Depending on the scattering length $a$, the lowest energy state may be a fundamental fermion (atom) or a bosonic molecule (dimer) or even a composite fermion formed from three atoms (trimer). Trimers are stable (lowest energy state) for an interval $(a^{(1)}_{c1})^{-1}<a^{-1}<(a^{(1)}_{c2})^{-1}$ which depends on the Yukawa or Feshbach coupling $\bar h$ related to the width of the Feshbach resonance. We have computed the binding energy $E_1<0$ of the trimer, cf. Fig. \ref{fig:Energies}. For $(a^{(2)}_{c1})^{-1}<a^{-1}<(a^{(2)}_{c2})^{-1}$ a second trimer bound state is present, with binding energy $E_2$. Close to the resonance, we find the (approximate) relations
\begin{eqnarray}
&&E_{n+1}=\exp\left( -\frac{2 \pi}{s_0} \right)E_n,\notag \\
&&(a^{(n+1)}_{c1,2})^{-1}=(a^{(n)}_{c1,2})^{-1}\exp\left( -\frac{\pi}{s_0} \right).
\label{eq:relations}
\end{eqnarray}
Our truncation yields $s_0=0.82$. This structure repeats for higher $n$, such that at the location of the Feshbach resonance for $a^{-1}\to 0$ an infinite tower of trimer bound states builds up. The relations, Eq. \eqref{eq:relations}, become exact for large $n$, if the quantity $s_0$ is computed precisely.

The origin of this repetition of self-similar features on different scales is due to a limit cycle scaling behavior of the renormalization flow. The same features appear periodically for $s_0 \ln\frac{k}{\Lambda}=-n \pi$, with $k$ the appropriate scale of the model. In other words, at the scale $k_{n+1}=k_n \exp\left(-\frac{\pi}{s_0}\right)$ one finds the same physics as at the scale $k_n$. This concerns all dimensionful quantities according to their dimension, e.g. $E_n\sim k_n^2$, $(a^{(n)}_{c1,2})^{-1}\sim k_n$. Also the three-body scattering length $\lambda_3$ depends periodically on the logarithm of the characteristic scale $k$, with an overall factor $k^{-4}$, cf. Eq. \eqref{eq:lambda3}.

We have only considered the vacuum in the present paper, but we can already infer some interesting features of what happens at nonzero density for the many-body ground state at $T=0$. The qualitative properties for small density follow from our computations by simple arguments of continuity. For small density, we find that a ``trion phase'' separates the BEC phase and the BCS phase. 

For increasing density, the chemical potential $\mu$ increases. The BEC phase occurs for small density for $a^{-1}>(a_{c2})^{-1}$. Due to the symmetries of the microscopic action, the effective potential for the bosonic field depends only on the $\text{SU(3)}\times\text{U(1)}$ invariant combination $\rho=\phi^\dagger\phi=\phi_1^*\phi_1+\phi_2^*\phi_2+\phi_3^*\phi_3$. It reads for small $\mu-\mu_0$, with $\mu_0$ the vacuum chemical potential,
\begin{equation}
U(\rho)=\frac{\lambda_\phi}{2}\rho^2-2(\mu-\mu_0)\rho,
\label{eq::effpot}
\end{equation}
where we use the fact that the term linear in $\rho$ vanishes for $\mu=\mu_0$, i.e. $m_\phi^2(\mu_0)=0$. The minimum of the potential shows a nonzero condensate
\begin{equation}
\rho_0=\frac{2(\mu-\mu_0)}{\lambda_\phi},\ U(\rho_0)=-\frac{2}{\lambda_\phi}(\mu-\mu_0)^2,
\end{equation}
with density
\begin{equation}
n=-\frac{\partial U(\rho_0)}{\partial \mu}=\frac{4}{\lambda_\phi}(\mu-\mu_0).
\end{equation}

For the BCS phase for $a<a_{c1}$ ($a_{c1}<0$), one has $\mu_0=0$. Nonzero density corresponds to positive $\mu-\mu_0$, and in this region the renormalization flow drives $m_\phi^2$ always to zero at some finite $k_c$, with BCS spontaneous symmetry breaking ($\rho_0>0$) induced by the flow for $k<k_c$. Both the BEC and BCS phases are therefore characterized by superfluidity with a nonzero expectation value of the boson field $\phi_0\neq0$ with $\rho_0=\phi_0^*\phi_0$. 

As an additional feature to the BCS-BEC crossover for a Fermi gas with two components, the expectation value for the bosonic field $\phi_0$ in the three component case also breaks the spin symmtry of the fermions (SU(3)). Due to the analogous in QCD this was called ``color superfluidity'' \cite{Honerkamp}. For any particular direction of $\phi_0$ a continuous symmetry $\text{SU(2)}\times\text{U(1)}$ remains. Acoording to the symmetry breaking $\text{SU(3)}\times\text{U(1)}\to \text{SU(2)}\times\text{U(1)}$, the effective potential has five flat directions.

For two identical fermions the BEC and BCS phases are not separated, since in the vacuum either $\mu_0=0$ or $m_\phi^2=0$. There is no phase transition, but rather a continuous crossover. For three identical fermions, however, we find a new \textit{trion phase} for $(a^{(1)}_{c1})^{-1}<a^{-1}<(a^{(1)}_{c2})^{-1}$. In this region the vacuum has $\mu_0<0$ and $m_\phi^2>0$. The atom fluctuations are cut off by the negative chemical potential and do not drive $m_\phi^2$ to zero, such that for small density $m_\phi^2$ remains positive. Adding a term $m_\phi^2 \rho$ to the effective potential \eqref{eq::effpot} we see that the minimum remains at $\rho_0=0$ as long as $m_\phi^2>2(\mu-\mu_0)$. No condensate of bosons occurs. The BEC and BCS phases that show both extended superfluidity through a spontaneous breaking of the $\text{SU(3)}\times\text{U(1)}$ symmetry, are now separated by a phase where $\phi_0=0$, such that the $\text{SU(3)}\times\text{U(1)}$ symmetry remains unbroken (or will be only partially broken).

Deep in the trion phase, e.g. for very small $|a^{-1}|$, the atoms and dimers can be neglected at low density since they both have a gap. The thermodynamics at low density and temperature is determined by a single species of fermions, the trions. In our approximation it is simply given by a noninteracting Fermi gas, with fermion mass $3M$ and chemical potenial $3(\mu-\mu_0)$. Beyond our approximation, we expect that trion interactions are induced by the fluctuations. While local trion interactions $\sim (\chi^*\chi)^2$ are forbidden by Fermi statistics, momentum dependent interactions are allowed. These may, however, be ``irrelevant interactions'' at low density, since also the relevant momenta are small such that momentum dependent interactions will be suppressed. Even if attractive interactions would induce a di-trion condensate, this has atom number six and would therefore leave a $\textrm{Z}_6$ subgroup of the U(1) transformations unbroken, in contrast to the BEC and BCS phase where only $\textrm{Z}_2$ remains. Furthermore, the trions are $\text{SU(3)}$-singlets such that the $\text{SU(3)}$ symmetry remains unbroken in the trion phase. The different symmetry properties between the possible condensates guarantee true quantum phase transitions in the vicinity of $a_{c1}$ and $a_{c2}$ for small density and $T=0$. We expect that this phase transition also extends to small nonzero temperature.

While deep in the trion phase the only relevant scales are given by the density and temperature, and possibly the trion interaction, the situation becomes more complex close to quantum phase transition points. For $a\approx a_{c1}$ we have to deal with a system of trions and atoms, while for $a\approx a_{c2}$ a system of trions and dimers becomes relevant. The physics of these phase transitions may be complex and rather interesting.

We finally comment on the precision of our computation. At nonzero density, our truncation may be improved by including in the two-body sector a four fermion coupling $\sim (\psi^{\dagger}\psi)^2$, which may be partially bosonized in favor of a running $\bar h$ \cite{FSDW}. Also the momentum dependence of the interactions $\sim \lambda_{\phi\psi}$ or $\lambda_\psi$ may be resolved beyond the pointlike approximation. Furthermore we expect an improvement from including a second atom-dimer scattering channel $\sim \lambda_{AD} \phi^\dagger\phi \psi^\dagger\psi$, cf. \cite{DKS}. These improvements are expected to change the quantitative values of $a_{c1}$, $a_{c2}$, $\mu_0$, and $s_0$, but not the qualitative situation. We believe that already the present truncation will yield a reliable picture of the qualitative properties of the phase diagram, once it is extended to nonzero density and temperature.

\end{document}